\documentclass[showpacs,preprintnumbers,superscriptaddress]{revtex4}
\pdfoutput=1
\usepackage{amsmath,amssymb,graphicx,bm}
\usepackage{epstopdf}
\usepackage{float}
\usepackage{subfig}
\usepackage{natbib}
\begin{document}
\title{Types of the geodesic motions in Kerr-Sen-AdS$_{4}$ spacetime}

\author{Ziqiang Cai}
\affiliation{College of Physics Science and Technology, Hebei University, Baoding 071002, China}

\author{Tong-Yu He}
\affiliation{College of Physics Science and Technology, Hebei University, Baoding 071002, China}

\author{Wen-Qian Wang}
\affiliation{College of Physics Science and Technology, Hebei University, Baoding 071002, China}

\author{Zhan-Wen Han}
\affiliation{College of Physics Science and Technology, Hebei University, Baoding 071002, China}
\affiliation{Yunnan Observatories, Chinese Academy of Sciences, Kunming 650216, China}

\author{Rong-Jia Yang \footnote{Corresponding author}}
\email{yangrongjia@tsinghua.org.cn}
\affiliation{College of Physics Science and Technology, Hebei University, Baoding 071002, China}
\affiliation{Hebei Key Lab of Optic-Electronic Information and Materials, Hebei University, Baoding 071002, China}
\affiliation{National-Local Joint Engineering Laboratory of New Energy Photoelectric Devices, Hebei University, Baoding 071002, China}
\affiliation{Key Laboratory of High-pricision Computation and Application of Quantum Field Theory of Hebei Province, Hebei University, Baoding 071002, China}

\begin{abstract}
We consider the geodesic motions in the Kerr-Sen-AdS$_4$ spacetime. We obtain the equations of motion for light rays and test particles. Using the parametric diagrams, we shown some regions where the radial and latitudinal geodesic motions are allowed. We analyse the impact of parameter related to dilatonic scalar on the orbit and find that it will result in more rich and complex orbital types.
\end{abstract}

\maketitle

\section{Introduction}
The Event Horizon Telescope has released the observed black hole shadow \cite{EventHorizonTelescope:2019dse} which allows chances to have a deeper understanding of gravitational field of massive object. Studying test particles and light rays in spacetimes has been a matter of interest for a long time: it is an important channel for understanding black holes and predicts a number of observational effects. The study of geodesic motion can be traced back to the early work done by Hagihara \cite{hagihara1930theory} who solved analytically the equations of motion of test particles and light rays in the Schwarzschild spacetime. It has been shown that the geodesic equations in Kerr, Reissner-Nordstr{\"o}m, and Kerr-Newman spacetimes have the same mathematical structure \cite{1983Chandrasekhar}. Since then, many works in the literatures have extensively investigated the equations of motion of particles and light rays in various spacetimes, see for example \cite{Hoseini:2016tvu, Soroushfar:2016esy,Hackmann:2008zza,Hackmann:2010zz, Soroushfar:2015wqa, Flathmann:2015xia}. The geodesic equations in some spacetimes can be analytically solved in terms of the Weierstrass fountions and the derivatives of Kleinian functions \cite{Hackmann:2010zz,Soroushfar:2016yea,Soroushfar:2016esy,Flathmann2016}. These methods have been applied to higher dimensional black holes \cite{Hackmann:2008tu, Kagramanova:2012hw, Diemer:2013fza,Diemer:2014lba}, to Taub-NUT and wormhole spacetime \cite{Kagramanova:2010bk, Diemer:2013hgn}, and to Kerr-Sen dilaton-axion black hole \cite{Soroushfar:2016yea}. Recently this analytical approach has been further developed and applied to the hyperelliptic case, where the analytical solutions of the equations of motion in the four-dimensional Schwarzschild-(A)dS, Reissner-Nordstr{\"o}m(A)dS, and Kerr-(A)dS spacetimes were presented \cite{Hackmann:2008zz,Hackmann:2008tu,Hackmann:2009nh,Hackmann:2010zz,Grunau:2010gd,Enolski:2010if}. The motions of test particles were also studied in various black string spacetimes \cite{Hackmann:2010ir,Hackmann:2009rp,Grunau:2013oca,Ozdemir:2004ne,Aliev:1988wv,Galtsov:1989ct,Chakraborty:1991mb}. Recently, Kerr geodesics in terms of Weierstrass elliptic functions are discussed in \cite{Cieslik:2023qdc}. Other works, see for example \cite{Grunau:2010gd,Kraniotis:2004cz,Kraniotis:2019ked,Hackmann:2010ir,Hackmann:2013pva,Garcia:2013zud,Hackmann:2014tga,Yang:2023snt}, also discussed the possible geodesic motions in various spacetime.

In \cite{Wu:2020cgf}, a solution including a nonzero negative cosmological constant into the Kerr-Sen solution was obtained. An analysis of all possible orbits for particles and light in the spacetime of this Kerr-Sen-AdS$_{4}$ black hole is still not presented. In this paper, we will fill this gap. We will consider the geodesic motion in the background of the Kerr-Sen-AdS$_{4}$ black hole and analyze in detail the possible orbit types.

The order of this paper is as follows. In Section II, we give a brief review of the Kerr-Sen-AdS$_4$ metric. In Section III, we present the equations for geodesic motions in the Kerr-Sen-AdS$_4$ spacetime. In Section IV, we give a full analysis of the geodetic equations. Finally, we will briefly summarise and discuss our results in Section V.

\section{The Kerr-Sen-AdS$_4$ black hole solution}
The Lagrangian including a nonzero negative cosmological constant into the four-dimensional gauged Einstein-Maxwell-dilaton-axion theory has the following form
\begin{equation}
\begin{aligned}
\mathcal{L}= \sqrt{-g}\left\{R-\frac{1}{2}\left(\partial \phi\right)^2-\frac{1}{2} e^{2 \phi}\left(\partial \chi\right)^2-e^{-\phi} F^2
+\frac{1}{l^2}\left[4+e^{-\phi}+e^{\phi}\left(1+\chi^2\right)\right]\right\}+\frac{\chi}{2} \varepsilon^{\mu \nu \rho \lambda} F_{\mu \nu}F_{\rho \lambda},
\end{aligned}
\end{equation}
where $g$ is the determinant of the metric, $R$ is the Ricci scalar, $\phi$ is the dilaton scalar field, $F_{\mu\nu}$ is the electromagnetic tensor and
$F^2=F_{\mu\nu}F^{\mu\nu}$, $\chi$ is the axion pseudoscalar field dual to the three-form antisymmetric tensor: $H=-e^{2\phi}\star d\chi$ and $H^2=H_{\mu\nu\sigma}H^{\mu\nu\sigma}$, $l$ is the cosmological scale, and $\varepsilon_{\mu\nu\rho\lambda}$ is the four-dimensional Levi-Civita antisymmetric tensor density. A solution for this Lagrangian, called Kerr-Sen-AdS$_4$ black hole, was obtained in \cite{Wu:2020cgf}. Written in terms of Boyer-Lindquist coordinates, it takes the  following form:
\begin{eqnarray}
\label{1}	
\mathrm{d}s^{2}=-\frac{\Delta_{r}}{\rho^{2}}\left(\mathrm{d}t-\frac{a\sin^{2}\theta }{\Xi}d\varphi\right)^{2}
+\frac{\rho^{2}}{\Delta_{r}}\mathrm{d}r^{2}+\frac{\rho^{2}}{\Delta_{\theta}}\mathrm{d}\theta^{2}+\frac{\Delta_{\theta}\sin^{2}\theta}{\rho^{2}}
\left(a\mathrm{d}t-\frac{r^{2}+2br+a^{2}}{\Xi}\mathrm{d}\varphi\right)^{2},
\end{eqnarray}
where
\begin{eqnarray}
\label{2}
\Delta_{r}=\left(1+\frac{r^{2}+2br}{l^{2}}\right)\left(r^{2}+2br+a^{2}\right)-2Mr,
\end{eqnarray}
\begin{eqnarray}
\label{3}
\Delta_{\theta}=1-\frac{a^{2}}{l^{2}}\cos^{2}\theta,~~~~\Xi=1-\frac{a^{2}}{l^{2}},~~~~\rho^{2}=r^{2}+2br+a^{2}\cos^{2}\theta,
\end{eqnarray}
in which $a=J/M$ is the angular momentum per unit mass of the black hole, $b=Q^{2}/2M$ is the dilatonic scalar charge, $M$ is the mass of the black hole, and $Q$ is the charge of the black hole. The horizons in metric (\ref{1}) are given by $\Delta_{r}=0$. The horizons are local at $\Delta_{r}=0$, meaning that there could be up to four horizons and one of them is probably a cosmological horizon. The contravariant metric components are give by
\begin{equation}
\begin{aligned}
\label{4}	g^{tt}=-\frac{(r^{2}+2br+a^{2})^{2}\Delta_{\theta}\sin^{2}\theta-a^{2}-a^{2}\Delta_{r}\sin^{4}\theta}{\rho^{2}\Delta_{\theta}\Delta_{r}\sin^{2}\theta},\\
g^{rr}=\frac{\Delta_{r}}{\rho^{2}},~~~~g^{\theta\theta}=\frac{\Delta_{\theta}}{\rho^{2}},~~~~g^{\varphi\varphi}=-\frac{(a^{2}\Delta_{\theta}\sin^{2}\theta-\Delta_{r})\Xi^{2}}{\rho^{2}\Delta_{\theta}\Delta_{r}\sin^{2}\theta},\\ g^{t\varphi}=g^{\varphi{t}}=\frac{(a\Delta_{r}\sin^{2}\theta-a\Delta_{\theta}(r^{2}+2br+a^{2})\sin^{2}\theta)\Xi}{\rho^{2}\Delta_{\theta}\Delta_{r}\sin^{2}\theta}.\\
\end{aligned}
\end{equation}
The Kerr-Sen-AdS$_4$ black hole (\ref{1}) reduces to the Kerr-AdS$_4$ solution \cite {Plebanski:1976gy,Carter:1968ks} for $b=0$ and reduces to the Kerr-Sen solution \cite{Sen:1992ua} when $l$ tends to infinity.

\section{The geodesic equations}
In this section, we will derive the equations of motion for Kerr-Sen-AdS$_4$ black hole (\ref{1}) by using the Hamilton-Jacobi formalism, and later we will introduce effective potentials for the $r$ and $\theta$ motion. The Hamilton-Jacobi equation is
\begin{eqnarray}
\label{5}
\frac{\partial{S}}{\partial{\tau}}+\frac{1}{2}g^{ij}\frac{\partial{S}}{\partial{x^{i}}}\frac{\partial{S}}{\partial{x^{j}}}=0,
\end{eqnarray}
which can be solved with an ansatz for the action
\begin{eqnarray}
\label{6}
S=\frac{1}{2}\varepsilon\tau-Et+L_{z}\phi+S_{\theta}(\theta)+S_{r}(r).
\end{eqnarray}
where the parameter $\varepsilon$ is equal to $1$ for particles and to $0$ for light, $\tau$ is an affine parameter along the geodesic. The energy $E$ and the angular momentum $L$, two constants of motion, are related to the the generalized momenta $P_{t}$ and $P_{\phi}$ as
\begin{eqnarray}
\label{7}
P_{t}=g_{tt}\dot{t}+g_{t\varphi}\dot{\varphi}=-E,~~~~
P_{\phi}=g_{\varphi\varphi}\dot{\varphi}+g_{t\varphi}\dot{t}=L,
\end{eqnarray}
where the dot denotes the derivative with respect to $\tau$. Since $g_{t\varphi}$ depends on $b$, this parameter will affect the energy $E$ and the angular momentum $L$ of the test particle, comparing with the case in Kerr-AdS$_4$ solution \cite{Hackmann:2009nh,Hackmann:2010zz}. Using Eqs. (\ref{5}), (\ref{6}), and (\ref{7}), we have
\begin{equation}
\begin{aligned}
\label{8}
\Delta_{\theta}\left(\frac{\partial{S}}{\partial{\theta}}\right)^{2}+\varepsilon{a^{2}}\cos^{2}\theta-\frac{2aEL\Xi-E^{2}a^{2}\sin^{2}\theta}{\Delta_{\theta}}
+\frac{L^{2}\Xi^{2}}{\Delta_{\theta}\sin^{2}\theta}\\
=-\Delta_{r}\left(\frac{\partial{S}}{\partial{r}}\right)^{2}-\varepsilon\left(r^{2}+2br\right)+\frac{\left(r^{2}+2br+a^{2}\right)^{2}E^{2}+a^{2}L^{2}\Xi^{2}
-2a\left(r^{2}+2br+a^{2}\right)EL\Xi}{\Delta_{r}}.\\
\end{aligned}
\end{equation}
The left-hand side of equation (\ref{8}) depends only on $\theta$ and the right-hand side depends only on $r$. With the ansatz Eq. (\ref{6}) and the Carter constant \cite{Carter:1968rr}, we obtain the equations of motion:
\begin{eqnarray}
\label{9}
\rho^{4}\left(\frac{\mathrm{d}r}{\mathrm{d}\tau}\right)^{2}=-\Delta_{r}\left[K+\varepsilon\left(r^{2}+2br\right)\right]+\left[\left(r^{2}+2br+a^{2}\right)E-aL\Xi\right]^{2},
\end{eqnarray}
\begin{eqnarray}
\label{10}
\rho^{4}\left(\frac{\mathrm{d}\theta}{\mathrm{d}\tau}\right)^{2}=\Delta_{\theta}\left(K-\varepsilon{a^{2}}\cos^{2}\theta\right)-\frac{1}{\sin^{2}\theta}
\left(aE\sin^{2}\theta-L\Xi\right)^{2},
\end{eqnarray}
\begin{eqnarray}
\label{11}	\rho^{2}\left(\frac{\mathrm{d}\varphi}{\mathrm{d}\tau}\right)=\frac{a\left(r^{2}+2br+a^{2}\right)E\Xi-a^{2}L\Xi^{2}}{\Delta_{r}}-\frac{1}{\Delta_{\theta}\sin^{2}\theta}
\left(aE\Xi\sin^{2}\theta-L\Xi^{2}\right),
\end{eqnarray}
\begin{eqnarray}
\label{12}
\rho^{2}\left(\frac{\mathrm{d}t}{\mathrm{d}\tau}\right)=\frac{E\left(r^{2}+2br+a^{2}\right)^{2}-a\left(r^{2}+2br+a^{2}\right)L\Xi}{\Delta_{r}}-\frac{\sin^{2}\theta}{\Delta_{\theta}}
\left(a^{2}E-\frac{aL\Xi}{\sin^{2}\theta}\right),
\end{eqnarray}
where $K$ is the Carter constant \cite{Carter:1968rr}.

The angular velocity is defined as: $\Omega=\dot{\phi}/\dot{t}$. For equatorial circular orbits, $\dot{r}=\dot{\theta}=\ddot{r}=0$, we have
\begin{eqnarray}
\begin{aligned}
\label{14}
\Omega &=\frac{-\partial_r g_{t \phi} \pm \sqrt{\left(\partial_r g_{t \phi}\right)^2-\left(\partial_r g_{t t}\right)\left(\partial_r g_{\phi \phi}\right)}}{\partial_r g_{\phi \phi}}\\
&=\frac{\rho ^2 \left(a^2-l^2\right)^2 \left[\frac{2 a \left((b+r) \left(a^2+2 r (2 b+r)\right)-l^2 M\right)}{\rho ^2 (a^2-l^2)}\pm 2 \sqrt{\frac{l^2 (b+r) \left(a^2 (b+r) \left(2 r (2 b+r)+l^2\right)+2 r (2 b+r) \left(r \left(4 b^2+l^2\right)+l^2 (b-M)+6 b r^2+2 r^3\right)\right)}{\rho ^4 \left(a^2-l^2\right)^2}}\right]}{2 l^2 \left[-a^4 (b+r)+a^2 \left(r \left(l^2-4 b^2\right)+l^2 (b+M)-6 b r^2-2 r^3\right)+2 l^2 r (b+r) (2 b+r)\right]},
\end{aligned}
\end{eqnarray}
where the +/- sign refers to corotating/counterrotating orbits, namely orbits with angular momentum parallel (antiparallel) to the spin of the central object. In this case, the energy $E$ and the angular momentum $L$, respectively, takes the form

\begin{eqnarray}
\begin{aligned}
E & =-\left(g_{t t}+\Omega g_{t \phi}\right) \dot{t} \\
&=\frac{\Delta _r \left[a^2+l^2 (a \Omega-1)\right]-a \Delta _{\theta} \left[a^3+a^2 \Omega l^2-a l^2+\Omega l^2 r (2 b+r)\right]}{\rho\sqrt{\Delta _r \left(a^2+l^2 (a \Omega-1)\right)^2-\Delta _{\theta} \left(a^3+a^2 \Omega l^2-a l^2+\Omega l^2 r (2 b+r)\right)^2}},
\end{aligned}
\end{eqnarray}
\begin{equation}
\begin{aligned}
L & =\left(g_{t \phi}+\Omega g_{\phi \phi}\right) \dot{t} \\
&=\frac{l^2 \left(\Delta _{\theta} \left(a^2+r (2 b+r)\right) \left(a^3+a^2 \Omega l^2-a l^2+\Omega l^2 r (2 b+r)\right)-a \Delta _r \left(a^2+l^2 (a \Omega-1)\right)\right)}{\rho ^2 (a-l)^2 (a+l)^2 \sqrt{\frac{\Delta _r \left(a^2+l^2 (a \Omega-1)\right)^2-\Delta _{\theta} \left(a^3+a^2 \Omega l^2-a l^2+\Omega l^2 r (2 b+r)\right)^2}{\rho ^2 (a-l)^2 (a+l)^2}}}.
\end{aligned}
\end{equation}

\begin{figure}
	\begin{minipage}{0.5\textwidth}
		\includegraphics[scale=0.7,angle=0]{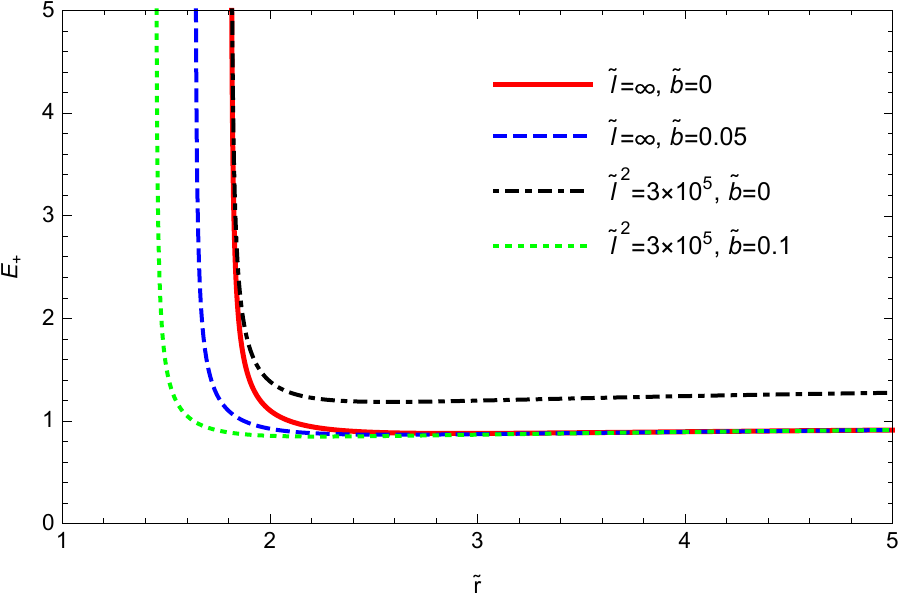}
	\end{minipage}%
	\begin{minipage}{0.5\textwidth}
		\includegraphics[scale=0.7,angle=0]{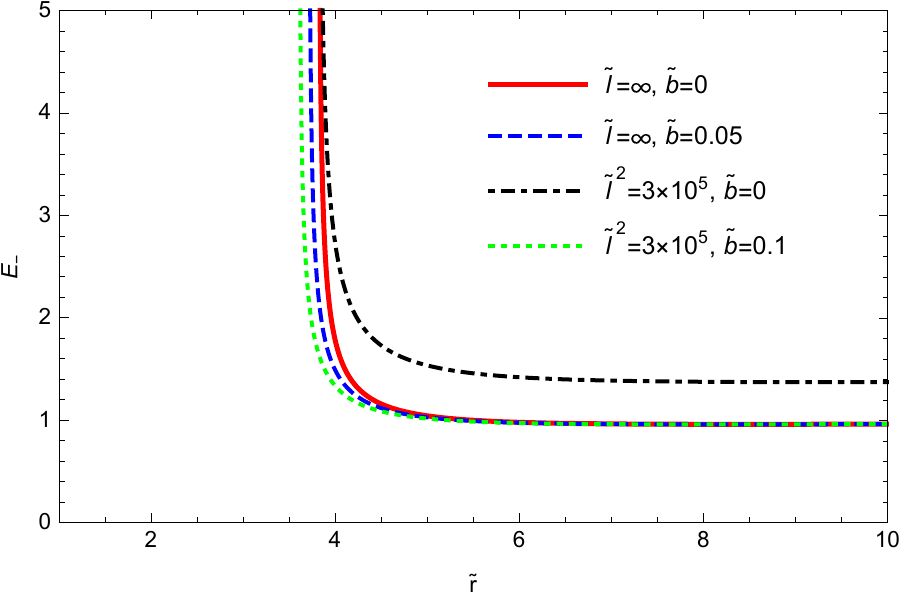}
	\end{minipage}
	\caption{Radial profiles of the functions $E_{+/-}(\tilde{r})$ for equatorial circular orbits with $\tilde{a}=0.8$ and various values of other parameters.}
\label{E}
\end{figure}

From Eqs. (\ref{9}) and (\ref{10}), we introduce two effective potentials $V_{\rm{reff}}$ and $V_{\theta\rm{eff}}$ such that $V_{\rm{reff}}=E$ and $V_{\theta\rm{eff}}=E$, corresponding to $\left(\frac{\mathrm{d}r}{\mathrm{d}\tau}\right)^{2}=0$ and $\left(\frac{\mathrm{d}\theta}{\mathrm{d}\tau}\right)^{2}=0$, respectively,
\begin{eqnarray}
\label{13}
V_{\rm{reff}}=\frac{aL\Xi\pm\sqrt{\Delta_{r}\left[K+\varepsilon(r^{2}+2br)\right]}}{r^{2}+2br+a^{2}},
\end{eqnarray}
\begin{eqnarray}
\label{14}
V_{\theta\rm{eff}}=\frac{L\Xi\pm\sqrt{\Delta_{\theta}(K-\varepsilon{a^{2}}\cos^{2}\theta)\sin^{2}\theta}}{a\sin^{2}\theta}.
\end{eqnarray}
To simplify the equations of motion, we adopt the Mino time $\lambda$ \cite{Mino:2003yg} connected to the proper time $\tau$ via $\frac{\mathrm{d}\tau}{\mathrm{d}\lambda}=\rho^{2}$, then the equations of motions can be rewritten as
\begin{eqnarray}
\label{15}
\left(\frac{\mathrm{d}r}{\mathrm{d}\lambda}\right)^{2}=-\Delta_{r}\left[K+\varepsilon\left(r^{2}+2br\right)\right]+\left[\left(r^{2}+2br+a^{2}\right)E-aL\Xi\right]^{2},
\end{eqnarray}
\begin{eqnarray}
\label{16}
\left(\frac{\mathrm{d}\theta}{\mathrm{d}\lambda}\right)^{2}=\Delta_{\theta}\left(K-\varepsilon{a^{2}}\cos^{2}\theta\right)
-\frac{1}{\sin^{2}\theta}\left(aE\sin^{2}\theta-L\Xi\right)^{2},
\end{eqnarray}
\begin{eqnarray}
\label{17}
\frac{\mathrm{d}\varphi}{\mathrm{d}\lambda}=\frac{a\left(r^{2}+2br+a^{2}\right)E\Xi-a^{2}L\Xi^{2}}{\Delta_{r}}
-\frac{1}{\Delta_{\theta}\sin^{2}\theta}\left(aE\Xi\sin^{2}\theta-L\Xi^{2}\right),
\end{eqnarray}
\begin{eqnarray}
\label{18}
\frac{\mathrm{d}t}{\mathrm{d}\lambda}=\frac{E\left(r^{2}+2br+a^{2}\right)^{2}-a\left(r^{2}+2br+a^{2}\right)L\Xi}{\Delta_{r}}
-\frac{\sin^{2}\theta}{\Delta_{\theta}}\left(a^{2}E-\frac{aL\Xi}{\sin^{2}\theta}\right).
\end{eqnarray}
Introducing some dimensionless quantities to rescale the parameters
\begin{eqnarray}
\label{19}
\tilde{r}=\frac{r}{M},~~\tilde{a}=\frac{a}{M},~~\tilde{t}=\frac{t}{M},~~\tilde{L}=\frac{L}{M},~~\tilde{l}=\frac{l}{M},
~~\tilde{b}=\frac{b}{M},~~\tilde{K}=\frac{K}{M^{2}},~~\gamma=M\lambda,
\end{eqnarray}
then the equations of motion (\ref{15})-(\ref{18}) can be formulated as
\begin{eqnarray}
\label{20}
\left(\frac{\mathrm{d}\tilde{r}}{\mathrm{d}\gamma}\right)^{2}=-\Delta_{\tilde{r}}\left[\tilde{K}+\varepsilon\left(\tilde{r}^{2}+2\tilde{b}\tilde{r}\right)\right]
+\left[\left(\tilde{r}^{2}+2\tilde{b}\tilde{r}+\tilde{a}^{2}\right)E-\tilde{a}\tilde{L}\Xi\right]^{2}=\tilde{R}(\tilde{r}),
\end{eqnarray}
\begin{eqnarray}
\label{21}
\left(\frac{\mathrm{d}\theta}{\mathrm{d}\gamma}\right)^{2}=\Delta_{\theta}\left(\tilde{K}-\varepsilon\tilde{a}^{2}\cos^{2}\theta\right)
-\frac{1}{\sin^{2}\theta}\left(\tilde{a}E\sin^{2}\theta-\tilde{L}\Xi\right)^{2}=\tilde{\Theta}(\theta),
\end{eqnarray}
\begin{eqnarray}
\label{22}
\frac{\mathrm{d}\varphi}{\mathrm{d}\gamma}=\frac{\tilde{a}\left(\tilde{r}^{2}+2\tilde{b}\tilde{r}+\tilde{a}^{2}\right)E\Xi
-\tilde{a}^{2}\tilde{L}\Xi^{2}}{\Delta_{\tilde{r}}}-\frac{1}{\Delta_{\theta}\sin^{2}\theta}\left(\tilde{a}E\Xi\sin^{2}\theta-\tilde{L}\Xi^{2}\right),
\end{eqnarray}
\begin{eqnarray}
\label{23}
\frac{\mathrm{d}\tilde{t}}{\mathrm{d}\gamma}=\frac{E\left(\tilde{r}^{2}+2\tilde{b}\tilde{r}+\tilde{a}^{2}\right)^{2}
-\tilde{a}\left(\tilde{r}^{2}+2\tilde{b}\tilde{r}+\tilde{a}^{2}\right)\tilde{L}\Xi}{\Delta_{\tilde{r}}}
-\frac{\sin^{2}\theta}{\Delta_{\theta}}\left(\tilde{a}^{2}E-\frac{\tilde{a}\tilde{L}\Xi}{\sin^{2}\theta}\right),
\end{eqnarray}
where
\begin{eqnarray}
\label{24}
\Delta_{\tilde{r}}=\left(1+\frac{\tilde{r}^{2}+2\tilde{b}\tilde{r}}{\tilde{l}^{2}}\right)\left(\tilde{r}^{2}+2\tilde{b}\tilde{r}+\tilde{a}^{2}\right)-2\tilde{r}.
\end{eqnarray}
And the effective potentials, the energy $E$, and the angular momentum $L$ can be expressed in terms of dimensionless quantity as
\begin{eqnarray}
\label{25}
\tilde{V}_{\rm{reff}}=\frac{\tilde{a}\tilde{L}\Xi\pm\sqrt{\Delta_{\tilde{r}}\left[\tilde{K}+\varepsilon
\left(\tilde{r}^{2}+2\tilde{b}\tilde{r}\right)\right]}}{\tilde{r}^{2}+2\tilde{b}\tilde{r}+\tilde{a}^{2}},
\end{eqnarray}
\begin{eqnarray}
\label{26}
\tilde{V}_{\theta\rm{eff}}=\frac{\tilde{L}\Xi\pm\sqrt{\Delta_{\theta}\left(\tilde{K}-\varepsilon \tilde{a}^{2}\cos^{2}\theta\right)\sin^{2}\theta}}{\tilde{a}\sin^{2}\theta}.
\end{eqnarray}

\begin{equation}
	\begin{aligned}
	\label{251}
	\Omega=\frac{\tilde{\rho }^2 \left(\tilde{a}^2-\tilde{l}^2\right)^2 \left[\frac{2 \tilde{a} \left(\left(\tilde{b}+\tilde{r}\right) \left(\tilde{a}^2+2 \tilde{r} \left(2 \tilde{b}+\tilde{r}\right)\right)-\tilde{l}^2\right)}{\tilde{\rho }^2 \left(\tilde{a}^2-\tilde{l}^2\right)}\pm 2 \sqrt{\frac{\tilde{l}^2 \left(\tilde{b}+\tilde{r}\right) \left(\tilde{a}^2 \left(\tilde{b}+\tilde{r}\right) \left(2 \tilde{r} \left(2 \tilde{b}+\tilde{r}\right)+\tilde{l}^2\right)+2 \tilde{r} \left(2 \tilde{b}+\tilde{r}\right) \left(\tilde{l}^2 \left(\tilde{b}+\tilde{r}-1\right)+2 \tilde{r} \left(\tilde{b}+\tilde{r}\right) \left(2 \tilde{b}+\tilde{r}\right)\right)\right)}{\tilde{\rho }^4 \left(\tilde{a}^2-\tilde{l}^2\right)^2}}\right]}{2 \tilde{l}^2 \left[\tilde{a}^2 \left(\tilde{l}^2 \left(\tilde{b}+\tilde{r}+1\right)-2 \tilde{r} \left(\tilde{b}+\tilde{r}\right) \left(2 \tilde{b}+\tilde{r}\right)\right)-\tilde{a}^4 \left(\tilde{b}+\tilde{r}\right)+2 \tilde{l}^2 \tilde{r} \left(\tilde{b}+\tilde{r}\right) \left(2 \tilde{b}+\tilde{r}\right)\right]},
	\end{aligned}
\end{equation}
\begin{eqnarray}
	\label{252}
	E=\frac{\Delta _{\tilde{r}} \left[\tilde{l}^2 \left(\Omega \tilde{a}-1\right)+\tilde{a}^2\right]-\tilde{a} \Delta _{\tilde{\theta}} \left[\Omega \tilde{a}^2 \tilde{l}^2-\tilde{a} \tilde{l}^2+\tilde{a}^3+\Omega \tilde{l}^2 \tilde{r} \left(2 \tilde{b}+\tilde{r}\right)\right]}{\tilde{\rho }^2 \left(\tilde{a}^2-\tilde{l}^2\right) \sqrt{\frac{\Delta _{\tilde{r}} \left[\tilde{l}^2 \left(\Omega \tilde{a}-1\right)+\tilde{a}^2\right]^2-\Delta _{\tilde{\theta}} \left[\Omega \tilde{a}^2 \tilde{l}^2-\tilde{a} \tilde{l}^2+\tilde{a}^3+\Omega \tilde{l}^2 \tilde{r} \left(2 \tilde{b}+\tilde{r}\right)\right]^2}{\tilde{\rho }^2 \left(\tilde{a}^2-\tilde{l}^2\right)^2}}},
\end{eqnarray}
\begin{eqnarray}
	\label{16}
	L=\frac{\tilde{l}^2 \left[\Delta _{\tilde{\theta}} \left(\tilde{a}^2+\tilde{r} \left(2 \tilde{b}+\tilde{r}\right)\right) \left(\Omega \tilde{a}^2 \tilde{l}^2-\tilde{a} \tilde{l}^2+\tilde{a}^3+\Omega \tilde{l}^2 \tilde{r} \left(2 \tilde{b}+\tilde{r}\right)\right)-\tilde{a} \Delta _{\tilde{r}} \left(\tilde{l}^2 \left(\Omega \tilde{a}-1\right)+\tilde{a}^2\right)\right]}{\tilde{\rho }^2 \left(\tilde{a}^2-\tilde{l}^2\right)^2 \sqrt{\frac{\Delta _{\tilde{r}} \left[\tilde{l}^2 \left(\Omega \tilde{a}-1\right)+\tilde{a}^2\right]^2-\Delta _{\tilde{\theta}} \left[\Omega \tilde{a}^2 \tilde{l}^2-\tilde{a} \tilde{l}^2+\tilde{a}^3+\Omega \tilde{l}^2 \tilde{r} \left(2 \tilde{b}+\tilde{r}\right)\right]^2}{\tilde{\rho }^2 \left(\tilde{a}^2-\tilde{l}^2\right)^2}}}.
\end{eqnarray}
Radial profiles of the functions $E_{+/-}(\tilde{r})$ ($F_{+/-}$ denotes that $\Omega$ takes $+$ in $F_+$ or $-$ in $F_-$) for $\tilde{a}=0.8$ and various values of other parameters are shown in the Fig. \ref{E}. The evolutions of $E_{+/-}$ for large $\tilde{r}$ tend to be consistent for particles circling around a Kerr ($\tilde{l}=\infty$ and $\tilde{b}= 0$), or a Kerr-dilation black hole ($\tilde{l}= \infty$ and $\tilde{b}\neq 0$), or a Kerr-Sen-AdS$_{4}$ ($\tilde{l}\neq \infty$ and $\tilde{b}\neq 0$), but the differences of $E_{+/-}$ in Kerr-AdS$_{4}$ ($\tilde{l}\neq \infty$ and $\tilde{b}= 0$) spacetime and in other three spacetime are obvious. For small $\tilde{r}$, the evolutions of $E_{+}$ in Kerr and Kerr-AdS$_{4}$ spacetime tend to be consistent, while distinguish from those in Kerr-dilation and  Kerr-Sen-AdS$_{4}$ spacetime; $E_{-}$ tend to be consistent in these four spacetime.

\begin{figure}
	\begin{minipage}{0.5\textwidth}
		\includegraphics[scale=0.7,angle=0]{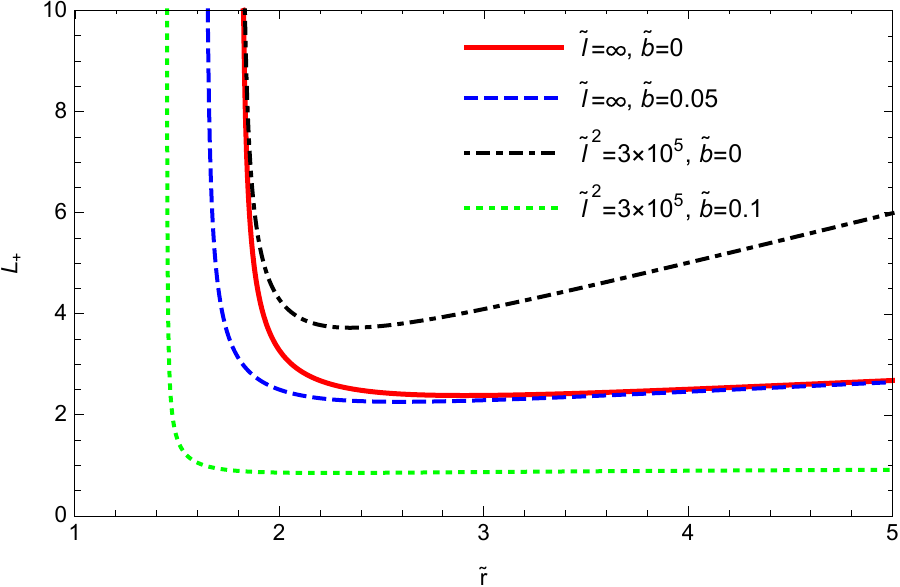}
	\end{minipage}%
	\begin{minipage}{0.5\textwidth}
		\includegraphics[scale=0.7,angle=0]{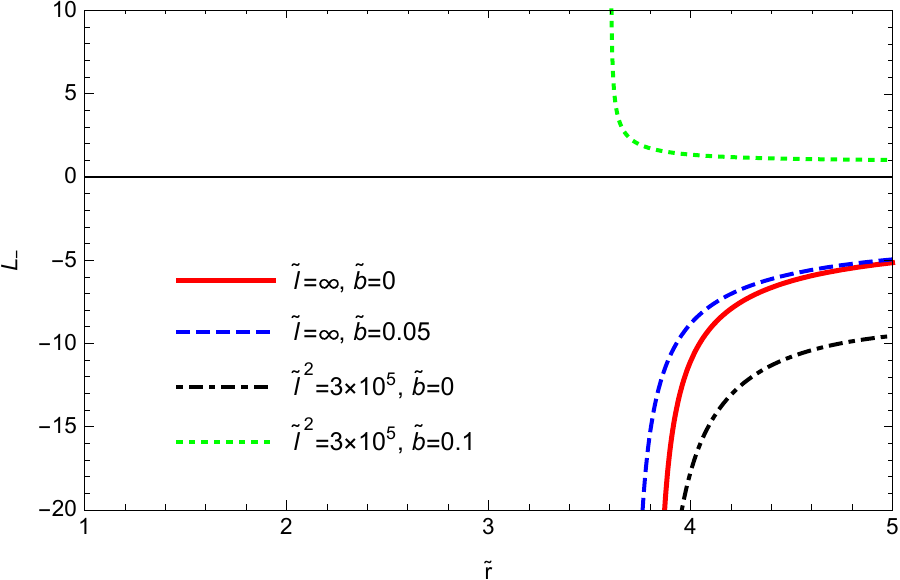}
	\end{minipage}
	\caption{Radial profiles of the functions $L_{+/-}(\tilde{r})$ for equatorial circular orbits with $\tilde{a}=0.8$ and various values of other parameters.}
\label{L}
\end{figure}

Radial profiles of the functions $L_{+/-}(\tilde{r})$ for $\tilde{a}=0.8$ and various values of other parameters are shown in the Fig. \ref{L}. The evolutions of $L_{+}$ for large $\tilde{r}$ tend to be consistent for particles circling around a Kerr or a Kerr-dilation black hole ($\tilde{l}= \infty$ and $\tilde{b}\neq 0$), but the differences of $L_{+}$ in Kerr-AdS$_{4}$ spacetime and in Kerr-Sen-AdS$_{4}$ spacetime are large, indicating that the parameter $\tilde{b}$ has a significant impact on $L_{+}$. For small $\tilde{r}$, the evolutions of $L_{+}$ in Kerr and Kerr-AdS$_{4}$ spacetime also tend to be consistent, while distinguish from those in Kerr-dilation and Kerr-Sen-AdS$_{4}$ spacetime. $L_{-}$ is positive in Kerr-Sen-AdS$_{4}$ spacetime while is negative in other three spacetimes.

\section{Analysis of the Geodesic equations}
In this section, we will give a full analysis of the geodesic equations of motion in the Kerr-Sen-AdS$_4$ spacetime and investigate the possible orbit types.

In \cite{Hackmann:2010zz}, two theorems were proofed for the case of Kerr-AdS$_4$ solution. We find those two theorems still holds for Kerr-Sen-AdS$_4$ solution \eqref{1}, though the parameter $b$ will change the values of the Carter constant $\tilde{K}$, the energy $\tilde{E}$ and the angular momentum $\tilde{L}$:

Theorem 1. The modified Carter constant $\tilde{Q}=\tilde{K}-\left(\tilde{a}E-\tilde{L}\Xi\right)^2$ is zero, if a geodesic lies entirely in the equatorial plane $\theta=\pi/2$ or if it hits the ring singularity $\rho=0$.

Theorem 2. All timelike and null geodesics have $\tilde{K}\geq 0$ if $1>\tilde{a}^2/\tilde{l}^2$. In this case $\tilde{K}=0$ implies $\tilde{Q}=0$ and the geodesic lies entirely in the equatorial plane.

The proof are similar to the case of Kerr-AdS$_4$ black hole in Einstein's gravity, for detail see \cite{Hackmann:2010zz}.

\subsection{Types of latitudinal motion}
First we consider the function $\tilde{\Theta}(\theta)$ in equation (\ref{21}). Let $v=\cos^{2}\theta$ with $v\in[0,1]$, the function $\tilde{\Theta}$ can be written as
\begin{eqnarray}
\label{27}
\tilde{\Theta}(v)=\left(1-\frac{\tilde{a}^{2}}{\tilde{l}^{2}}v\right)\left(\tilde{K}-\varepsilon\tilde{a}^{2}v\right)-\frac{1}{1-v}\left[\tilde{a}E\left(1-v\right)-\tilde{L}\Xi\right]^{2}.
\end{eqnarray}
Geodesic motion is possible only for $\tilde{\Theta}(v)\geq 0$, which also implies that $\tilde{K}\geq 0$ in all spacetimes with $1>\tilde{a}^2/\tilde{l}^2$. The zeros of $\tilde{\Theta}(\theta)$ are the turning points of the latitudinal motion. Assuming that $\tilde{\Theta}(v)$ has some zeros in $[0,1]$, the number of
zeros changes only if: (i) a zero crosses $0$ or $1$, or (ii) a double zero occurs. If $v=0$ is a zero, then
\begin{eqnarray}
\label{28}
\tilde{\Theta}(v=0)=\tilde{K}-\left(\tilde{a}E-\tilde{L}\Xi\right)^{2},
\end{eqnarray}
or
\begin{eqnarray}
\label{29}
\tilde{L}=\frac{\tilde{a}E\pm\sqrt{\tilde{K}}}{\Xi}.
\end{eqnarray}
From Eq. (\ref{27}), we see that $v=1$ is a pole of $\tilde{\Theta}(v)$ for $\tilde{L}\neq0$. So $v=1$ is a zero of $\tilde{\Theta}(v)$ only if $\tilde{L}=0$, therefore we have
\begin{eqnarray}
\label{30}
\tilde{\Theta}(v=1,\tilde{L}=0)=\left(1-\frac{\tilde{a}^{2}}{\tilde{l}^{2}}\right)\left(\tilde{K}-\varepsilon\tilde{a}^{2}\right).
\end{eqnarray}
So for $\tilde{\Theta}(v=1,\tilde{L}=0)=0$, we have $\tilde{K}=\varepsilon\tilde{a}^{2}$. In order to remove the pole of $\tilde{\Theta}(v)$ at $v=1$, we consider another function
\begin{eqnarray}
\label{31}
\tilde{\Theta}'(v)=(1-v)\left(1-\frac{\tilde{a}^{2}}{\tilde{l}^{2}}v\right)\left(\tilde{K}-\varepsilon\tilde{a}^{2}v\right)-\left[\tilde{a}E\left(1-v\right)-\tilde{L}\Xi\right]^{2},
\end{eqnarray}
where $\tilde{\Theta}(v)=\frac{1}{1-v}\tilde{\Theta}'(v)$. The double zeros satisfy the following conditions,
\begin{eqnarray}
\label{32}
\tilde{\Theta}'(v)=0 \qquad {\rm{and}} \qquad \frac{\mathrm{d}\tilde{\Theta}'(v)}{\mathrm{d}v}=0,
\end{eqnarray}
which implies
\begin{eqnarray}
\label{33}
\tilde{L}=\frac{\left(6E\pm\sqrt{36E^{2}-\frac{36}{\tilde{l}^{2}}\tilde{K}}\right)\left(-\frac{12}{\tilde{l}^{2}}\tilde{a}^{2}+12\right)}{-\frac{144}{\tilde{l}^{2}}\tilde{a}\Xi}.
\end{eqnarray}

We can use these informations to analyse the $\theta$ motion of all possible geodesics for given parameters of the black hole, $\tilde{a}$, $\tilde{b}$, and $\tilde{l}$. We see that Eq. (\ref{27}) doesn't depend obviously on the parameter $b$, but it will change the number of zeros or the positions of the zeros via changing the Carter constant $\tilde{K}$, the energy $\tilde{E}$ and the angular momentum $\tilde{L}$, comparing with the case for Kerr-AdS$_4$ solution in Einstein's gravity \cite{Hackmann:2010zz, Hackmann:2009nh} or the case for Kerr-AdS$_4$ solution with $q=0$ in $f(R)$ gravity \cite{Soroushfar:2016esy}.

\begin{figure}
\centering
\includegraphics[height=10cm,width=10cm]{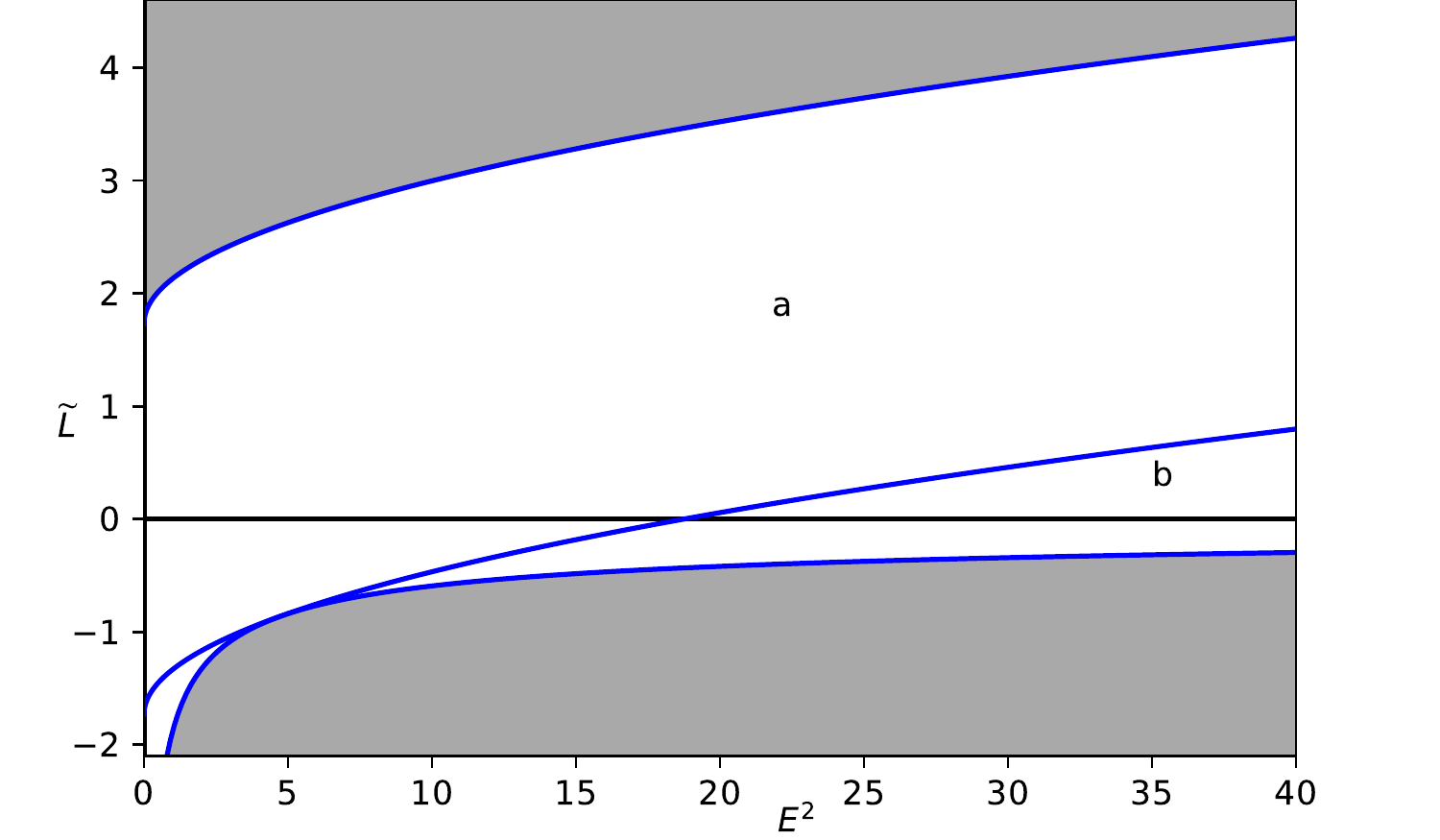}
\caption{Parametric $\tilde{L}-E^{2}$-diagram for the function $\tilde{\Theta}$ with $\varepsilon=1, \tilde{a}=0.4, \tilde{K}=3, \tilde{l}^{2}=3\times10^{5}$. $\tilde{\Theta}$ possesses one zero in region a and two zeros in region b. In the grey areas, geodesic motion is not allowed.}
\label{A}
\end{figure}

We plot parametric $\tilde{L}-E^{2}$ diagrams in Fig. \ref{A} from the condition of $v=0$ being a zero (Eq. (\ref{29})) and the condition of double zeros (Eq. (\ref{33})). The half plane is divided into four regions by the curves. The boundaries of region a are given by $\tilde{L}=\frac{\tilde{a}E\pm\sqrt{\tilde{K}}}{\Xi}$, it will get lager if $\tilde{K}$ grows, or it will shift up or down if $\tilde{a}$ changes. In regions a and b, geodesic motions are possible, because in all other regions $\Theta(v)$ is negative for all $v\in(0,1)$. The function $\tilde{\Theta}$ has a single zero in region a, where the geodesics will cross the equatorial plane ($\tilde{K}>(\tilde{a}E-\tilde{L}\Xi)^{2}$ or $\tilde{Q}>0$). In region b, the function $\tilde{\Theta}$ has two zeros, corresponding to motion above or below the equatorial plane ($\tilde{K}<(\tilde{a}E-\tilde{L}\Xi)^{2}$ or $\tilde{Q}<0$). If $\tilde{K}=(\tilde{a}E-\tilde{L}\Xi)^{2}$, the geodesics will remain in the equatorial plane.

\subsection{Types of radial motion}
A radial geodesic motion is possible if $\tilde{R}(\tilde{r})\geq 0$. The zeros of the function $\tilde{R}$ in Eq. (\ref{20}) are the turning points of orbits of particles and lights, so $\tilde{R}(\tilde{r})=0$ determines the possible types of orbits. Since the polynomial $\tilde{R}(\tilde{r})$ is of degree six in $\tilde{r}$ for massive particle, it has in general six possibly complex zeros of which the real zeros are of interest for the type of motion. The parameter $b$ will change the number of zeros or the positions of the zeros for other parameters taking the same values in the Kerr-AdS$_4$ case \cite{Hackmann:2009nh,Hackmann:2010zz}. If $\tilde{r}=0$ is an allowed value of $\tilde{r}(\gamma)$, we have
\begin{eqnarray}
	\label{R0} 0\leq\tilde{R}(0)=-\tilde{a}^2\left[\tilde{K}-\left(\tilde{a}E-\tilde{L}\Xi\right)^2\right]=-\tilde{a}^2\tilde{Q}.
\end{eqnarray}
Therefor $\tilde{r}=0$ can only be crossed if $\tilde{Q}\leq 0$, which corresponds to region b of the $\theta$ motion. Since $\tilde{Q}>0$ in region a of the $\theta$ motion, a transition from positive to negative $\tilde{r}$ is not possible. The number of real zeros of $\tilde{R}$ changes if a double zero occurs:
\begin{eqnarray}
	\label{34}
	\tilde{R}(\tilde{r})=0 \qquad  {\rm{and}}  \qquad \frac{\mathrm{d}\tilde{R}(\tilde{r})}{\mathrm{d}\tilde{r}}=0.
\end{eqnarray}
When the number of real zeros $\tilde{R}$ changes, the type of orbits will change.

Now, we discuss the possible types of orbit on which $\tilde{R}(\tilde{r})\geq 0$. The types of orbit were discussed in detail in \cite{Hackmann:2010zz, Soroushfar:2016esy}. Let $r_1$ and $r_2$ denote the roots of $\tilde{R}(\tilde{r})$, $\tilde{r}_+$ be the outer event horizon and $\tilde{r}_-$ be the inner event horizon.

1. Transit orbit (TrO): $-\infty<\tilde{r}<\infty$. The particle starts from $\pm\infty$ and goes to $\mp\infty$.

2. Escape orbit (EO): $r_1\leq\tilde{r}<\infty$ with $r_1> \tilde{r}_+$, or $-\infty<\tilde{r}\leq r_1$ with $r_1<\tilde{r}_-$. The particle approaches the black hole but turns around at a certain point to escape towards infinity.

3. Two-world escape orbit (TEO): $r_1\leq \tilde{r}<\infty$ with $r_1<\tilde{r}_-$, or $-\infty<\tilde{r}\leq r_1$ with $r_1>\tilde{r}_+$. The particle crosses the horizon twice and can enter another universe.

4. Crossover one-world escape orbit (COEO): $\tilde{r}_-<r_1<\tilde{r}_+$ with $r_1\leq\tilde{r}<\infty$ or $-\infty<\tilde{r}\leq r_1$. The particle crosses the outer horizon or inner horizon and can enter another universe.

5. Bound orbit (BO): $r_1\leq \tilde{r}\leq r_2$ with $r_1, r_2>\tilde{r}_+$ or $r_1, r_2< \tilde{r}_-$ or $\tilde{r}_-<r_1, r_2<\tilde{r}_+$. The particle oscillates between $r_1$ and $r_2$.

6. Crossover one-world bound orbit (COBO): $r_1\leq \tilde{r}\leq r_2$ with $r_1<\tilde{r}_-$ and $\tilde{r}_-<r_2< \tilde{r}_+$, or $r_1\leq \tilde{r}\leq r_2$ with $\tilde{r}_-<r_1<\tilde{r}_+$ and $r_2> \tilde{r}_+$. The particle oscillates between $r_1$ and $r_2$, crosses the inner or outer horizon.

7. Many-world bound orbit (MBO): $r_1\leq \tilde{r}\leq r_2$ with  $r_2>\tilde{r}_+$ and $r_1< \tilde{r}_-$.  The particle crosses the horizon multiple times and can enter another universe.

8. Circular orbit (CO): $\tilde{R}(\tilde{r})$ has real double roots. The particle circles around the black hole with $r_1=r_2>\tilde{r}_+$ (COO). The particle circles outside the inner horizon with $r_1=r_2< \tilde{r}_-$ (COI). the particle circles in the black hole with $\tilde{r}_-<r_1=r_2<\tilde{r}_+$ (COIn).

In the following, we will discuss in detail the types of orbit in two cases: $\tilde{R}(\tilde{r})$ has no real zero or has two real zeros. Other cases can be discussed similarly, but are much more complex. Because the more zeros, the more types of orbits there are.

Region I: $\tilde{R}$ has no real zero. If $\tilde{Q}<0$, we have $\tilde{R}(\tilde{r})> 0$ with $\tilde{r}\in (-\infty, +\infty)$, orbit types: TrO. If $\tilde{Q}=0$, we have $\tilde{r}\in [0, +\infty)$, orbit types: TEO.

Region II: $\tilde{R}$ has two real zeros. There are two cases needed to consider: these two zeros are double zeros or they are not.

Case A: these two zeros are not double zeros.

(1) If $\tilde{Q}\geq 0$: (a) $r_1\leq \tilde{r}\leq r_2$ with $r_1, r_2>\tilde{r}_+$ or $r_1, r_2< \tilde{r}_-$ or $\tilde{r}_-<r_1, r_2<\tilde{r}_+$, orbit type: BO; (b) $r_1\leq \tilde{r}\leq r_2$ with $r_1<\tilde{r}_-$ and $\tilde{r}_-<r_2< \tilde{r}_+$, or with $\tilde{r}_-<r_1<\tilde{r}_+$ and $r_2> \tilde{r}_+$, orbit type: COBO; (c) $r_1\leq \tilde{r}\leq r_2$ with $r_2>\tilde{r}_+$ and $r_1< \tilde{r}_-$, orbit type: MBO; (d) $\tilde{r}\leq r_1$ with $r_1<0$ or $\tilde{r}\geq r_2$ with $r_2< \tilde{r}_-$, orbit type: EO or TEO; (e) $\tilde{r}\leq r_1$ with $r_1<0$ or $\tilde{r}\geq r_2$ with $\tilde{r}_-< r_2<\tilde{r}_+$, orbit type: EO or COEO; (f) $\tilde{r}\leq r_1$ with $r_1<0$ or $\tilde{r}\geq r_2$ with $r_2>\tilde{r}_+$, orbit type: EO.

(2) If $\tilde{Q}\leq 0$: (a) $r_1\leq \tilde{r}\leq r_2$ with $r_1\leq 0$ and $r_2< \tilde{r}_-$, orbit type: BO; (b) $r_1\leq \tilde{r}\leq r_2$ with $r_1\leq0$ and $\tilde{r}_-<r_2< \tilde{r}_+$, orbit type: COBO; (c) $r_1\leq \tilde{r}\leq r_2$ with  $r_2>\tilde{r}_+$ and $r_1\leq 0$, orbit type: MBO; (d) $\tilde{r}\leq r_1$ or $\tilde{r}\geq r_2$ with $r_1<r_2<0$, orbit type: EO or TEO; (e) $\tilde{r}\leq r_1$ or $\tilde{r}\geq r_2$ with $0\leq r_1< r_2<\tilde{r}_-$, orbit type: EO or TEO; (f) $\tilde{r}\leq r_1$ or $\tilde{r}\geq r_2$ with $0\leq r_1<\tilde{r}_-<r_2<\tilde{r}_+$, orbit type: EO or COEO; (g) $\tilde{r}\leq r_1$ or $\tilde{r}\geq r_2$ with $\tilde{r}_-< r_1<r_2<\tilde{r}_+$, orbit type: COEO; (h) $\tilde{r}\leq r_1$ or $\tilde{r}\geq r_2$ with $0\leq r_1<\tilde{r}_-<\tilde{r}_+<r_2$, orbit type: EO; (i) $\tilde{r}\leq r_1$ or $\tilde{r}\geq r_2$ with $\tilde{r}_-< r_1<\tilde{r}_+<r_2$, orbit type: COEO or EO; (j) $\tilde{r}\leq r_1$ or $\tilde{r}\geq r_2$ with $\tilde{r}_+<r_1 <r_2$, orbit type: TEO or EO.

Case B: those two zeros are double zeros.

If $r_1=r_2>\tilde{r}_+$, orbit type: COO; if $r_1=r_2< \tilde{r}_-$, orbit type: COI; if $\tilde{r}_-<r_1=r_2<\tilde{r}_+$, orbit type: COIn.

Region III: $\tilde{R}$ has four real zeros and $\tilde{R}(\tilde{r})\geq 0$. Possible orbit types: EO, TEO, CO, BO, COBO, MBO, COEO.

Region IV: all six zeros of $\tilde{R}$ are  real and $\tilde{R}(\tilde{r})\geq 0$. Possible orbit types: EO, TEO, CO, BO, COBO, MBO, COEO.

\begin{figure}
	\begin{minipage}{0.5\textwidth}
		\includegraphics[scale=0.5,angle=0]{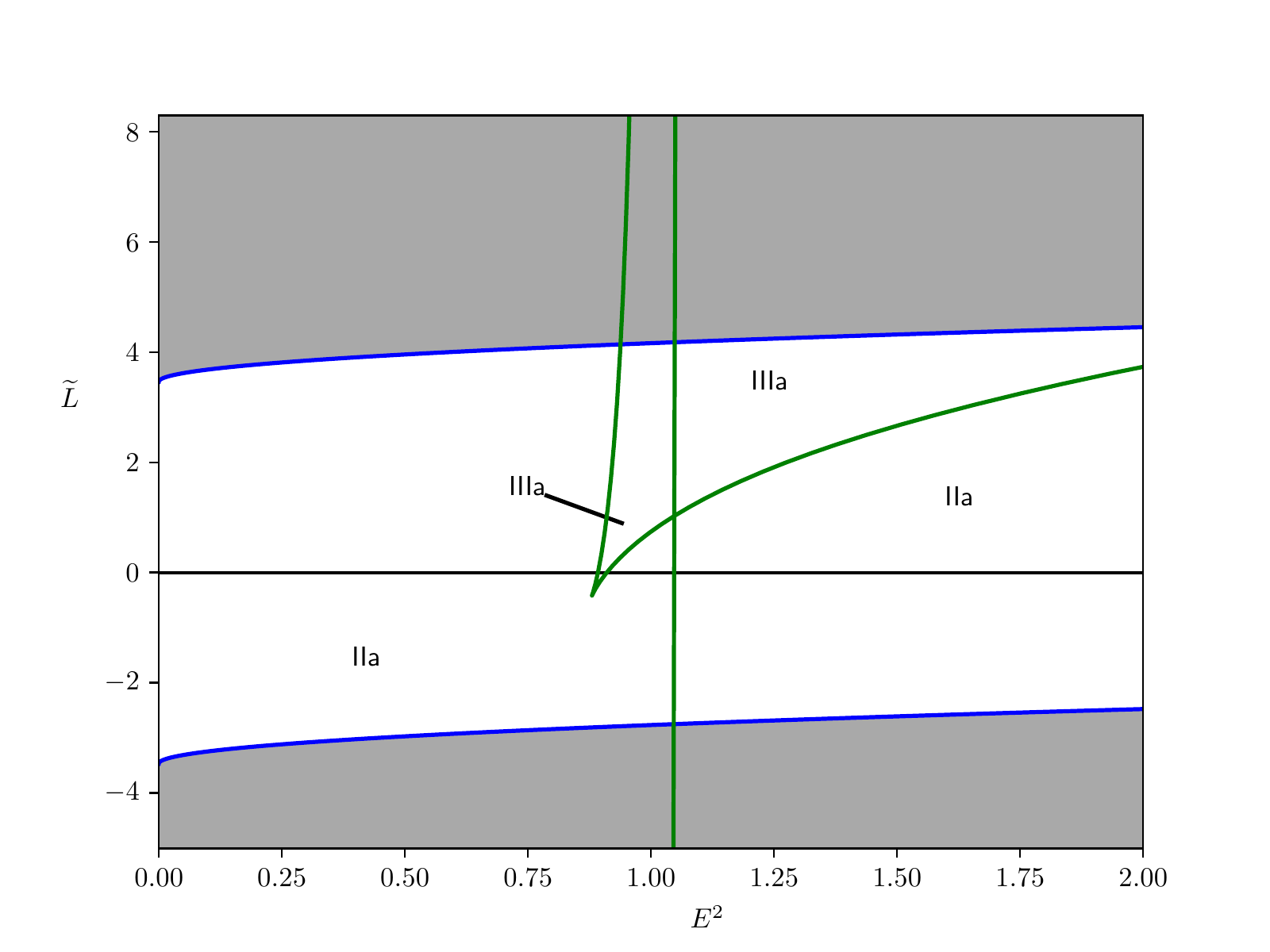}
	\end{minipage}%
	\begin{minipage}{0.5\textwidth}
		\includegraphics[scale=0.5,angle=0]{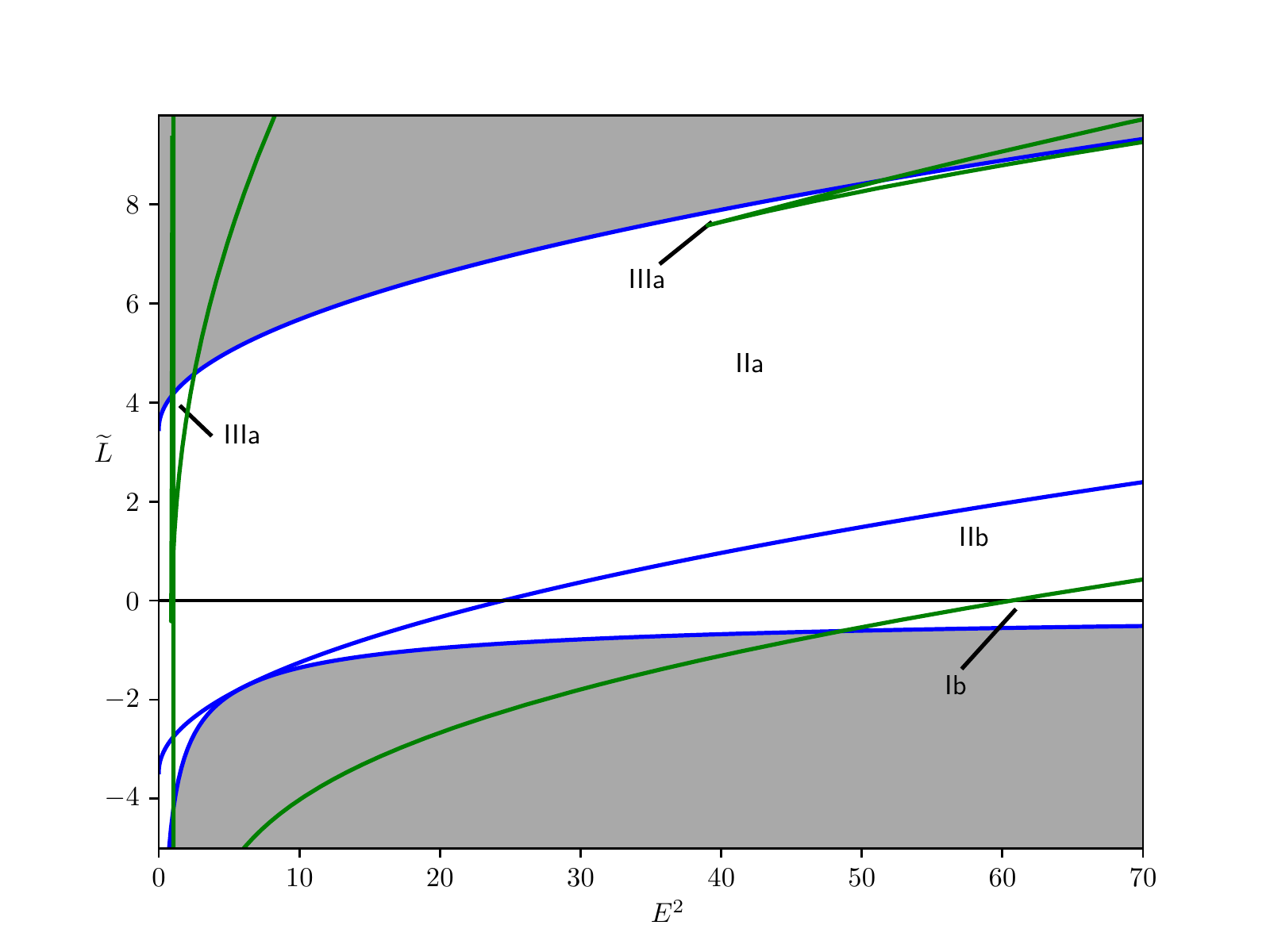}
	\end{minipage}
	\caption{Combined $\tilde{L}-E^{2}$ diagrams of the $\tilde{r}$ motion (green lines) and $\theta$ motion (blue lines) with $\varepsilon=1$, $\tilde{a}=0.7$, $\tilde{K}=12$,
$\tilde{b}=0.175$, $\tilde{l}^{2}=3\times10^{5}$. In regions marked with ``b", $\tilde{r}=0$ can be crossed but $\theta=\pi/2$ is never crossed.}
\label{el1}
\end{figure}

From the condition of double zero we can plot parametric $\tilde{L}-E^{2}$ diagrams, see, for example, in Fig. \ref{el1}. The polynomial $\tilde{R}$ has 2 positive zeros in left region IIa, 1 negative and 1 positive zeros in right region IIa, 4 positive in left region IIIa, 3 positive and 1 negative zeros in right region IIIa for the left figure; no zeros in region Ib, 1 negative and 1 positive zeros in region IIa, 2 negative zeros in region IIb, 4 positive in left region IIIa, 3 positive and 1 negative zeros in right region IIIa for the right figure. In regions marked with the letter ``a", the orbits cross $\theta=\pi/2$ but not $\tilde{r}=0$. Whereas in regions marked with the letter ``b", $\tilde{r}=0$ can be crossed but $\theta=\pi/2$ is never crossed. The $\theta$ equation dose not allow geodesic motion in the grey areas. The $\tilde{L}-E^{2}$ diagrams for Kerr, Kerr-AdS$_4$, and Kerr-dilaton spacetime are also presented for compare, there are 4 positive zeros on the left side of the vertical line while 3 positive and 1 negative zeros on the other side in region IIIa.

\begin{figure}
	\begin{minipage}{0.33\textwidth}
		\includegraphics[scale=0.33,angle=0]{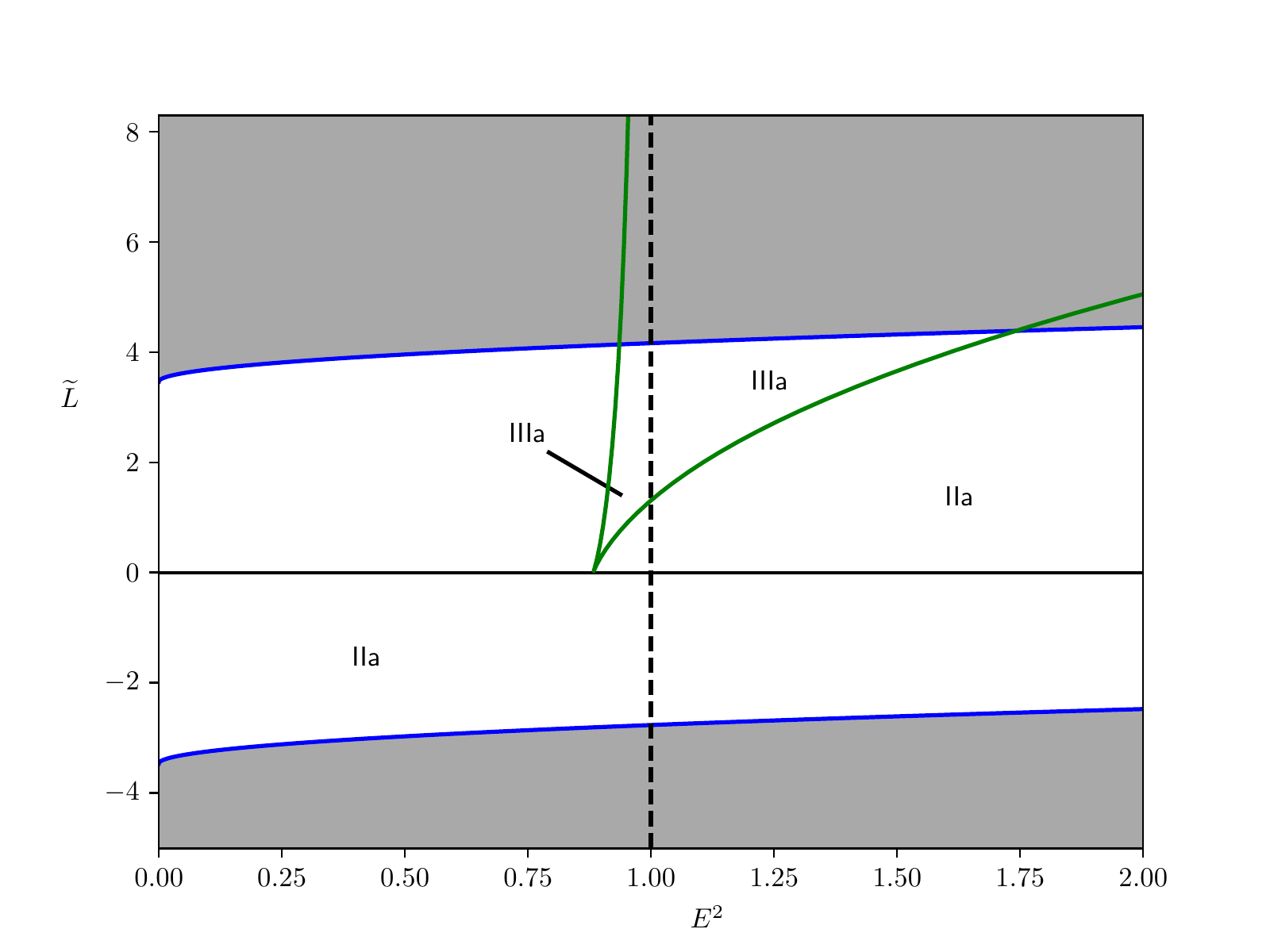}
	\end{minipage}%
	\begin{minipage}{0.33\textwidth}
		\includegraphics[scale=0.33,angle=0]{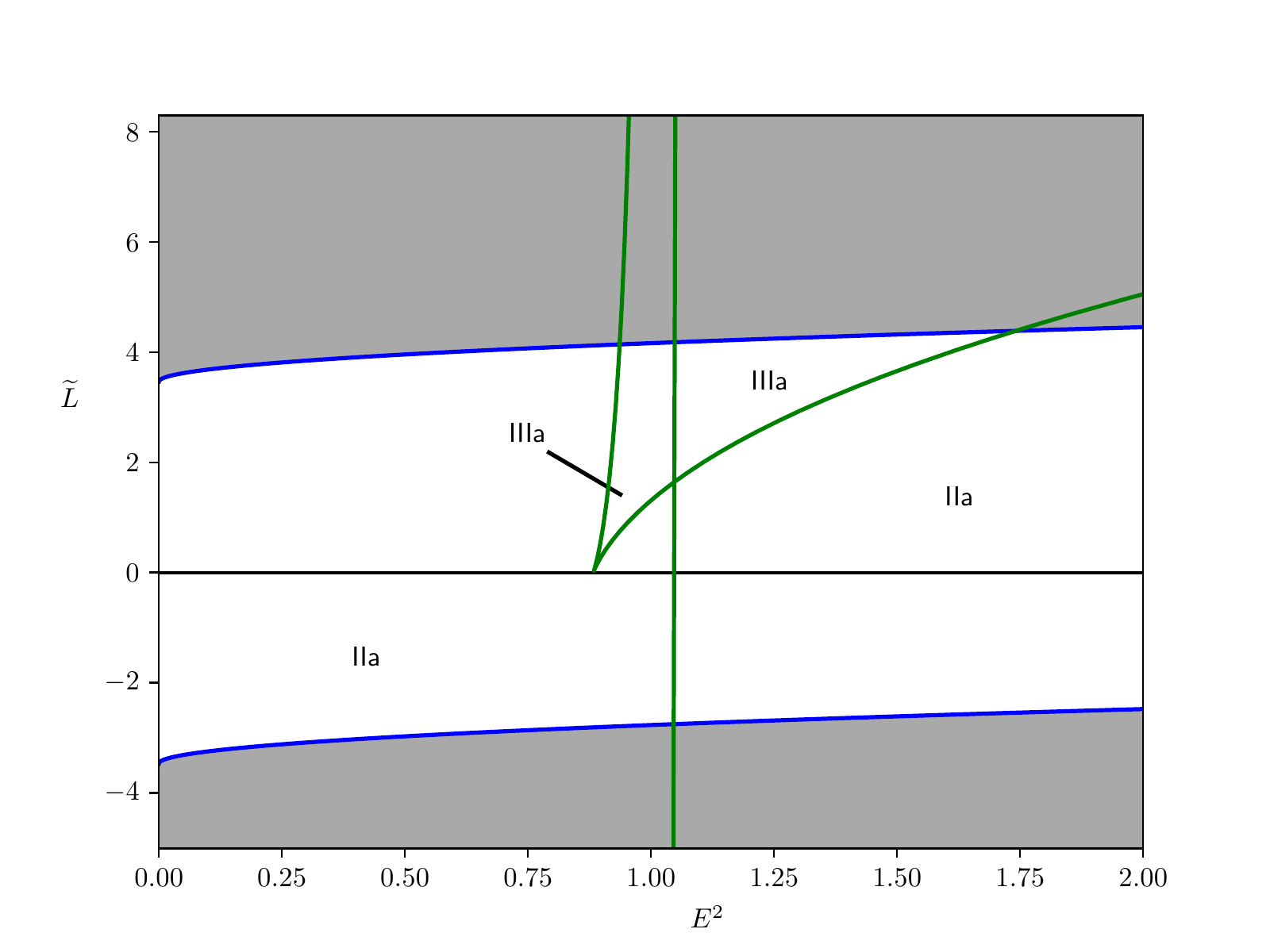}
	\end{minipage}
\begin{minipage}{0.33\textwidth}
		\includegraphics[scale=0.33,angle=0]{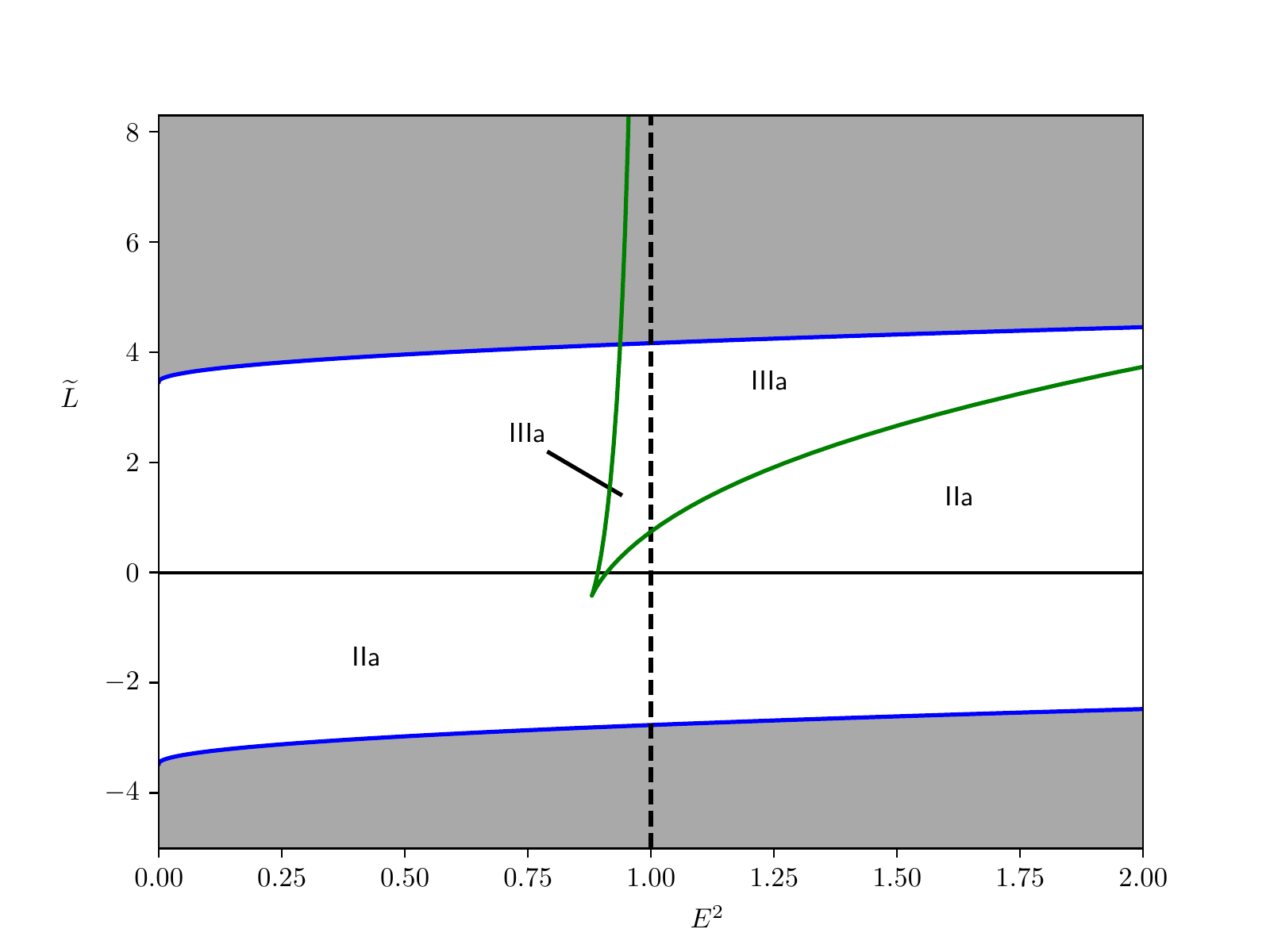}
	\end{minipage}
	\caption{Combined $\tilde{L}-E^{2}$ diagrams of the $\tilde{r}$ motion (green lines) and $\theta$ motion (blue lines) with $\varepsilon=1$, $\tilde{a}=0.7$, $\tilde{K}=12$,
and $\tilde{b}=0$, $\tilde{l}^{2}=\infty$ in left column; $\tilde{b}=0$, $\tilde{l}^{2}=3\times10^{5}$ in middle column; $\tilde{b}=0.175$, $\tilde{l}^{2}=\infty$ in right colum. In regions marked with ``a", the orbits cross $\theta=\pi/2$ but not $\tilde{r}=0$. The $\theta$ equation dose not allow geodesic motion in the grey areas.}
\label{el}
\end{figure}

As shown in \cite{Hackmann:2010zz,Hackmann:2008zz,Hackmann:2008zza}, a non-vanishing cosmological constant can dramatically change the possible structure of orbits and $\tilde{L}-E^2$ diagram. See for example: comparing with $\Lambda=0$ case, region in $\tilde{L}-E^2$ with four real zeros of $\tilde{R}(\tilde{r})$ becomes larger for $\Lambda<0$; and the transit orbit in region where $\tilde{R}(\tilde{r})$ has no real zero is transformed to a bound orbit in region where $\tilde{R}(\tilde{r})$ has two real zeros for $\Lambda<0$. However, these might not necessarily be the cases here, because parameter $b$ will affect the evolution of $\tilde{R}(\tilde{r})$. See in Figs. \ref{el} and \ref{el1}, region III becomes lager when parameter $\tilde{b}$ is nonzero, while a vertical line switch from lim$_{\tilde{r}\rightarrow\infty}\tilde{R}(\tilde{r})=\infty$ to lim$_{\tilde{r}\rightarrow\infty}\tilde{R}(\tilde{r})=-\infty$ appears when the cosmological constant is nonzero.

We observe from the discussions above that the parameter $b$ will result in rich and complex orbital types, comparing to the case of Kerr-AdS$_4$ \cite{Hackmann:2010zz}. To be more specific about this point, we consider an interesting case: $\tilde{r}=0$ as a double zero of $\tilde{R}$. From $\tilde{R}(0)=0$, yields
\begin{equation}
\tilde{a}=0,~~~~{\rm{or}}~~~~ \tilde{K}=\left(\tilde{a}E-\tilde{L}\Xi\right)^2.
\end{equation}
From $\frac{d\tilde{R}}{d\tilde{r}}(0)=0$, we have
\begin{equation}
\tilde{b}=1,~~~~{\rm{or}}~~~~\tilde{b}=\frac {\left( E\tilde{a}-\tilde{L}\Xi \right) ^{2}{\tilde{l}}^{2}}{{E}^{2}{\tilde{a}}^{4}-{E}^{2}{\tilde{a}}
^{2}{\tilde{l}}^{2}-2\,E\tilde{L}\Xi{\tilde{a}}^{3}+{\tilde{L}}^{2}{\Xi}^{2}{\tilde{a}}^{2}+{\tilde{L}}^{2}{\Xi}^{2}{\tilde{l}}^{2}+
{\tilde{a}}^{2}{\tilde{l}}^{2}}.
\end{equation}
We find that $\tilde{r}=0$ is not a double zero of $\tilde{R}$ in the case of Kerr-AdS$_4$ with $\tilde{a}=0$ \cite{Hackmann:2010zz}, but it is a double zero of $\tilde{R}$ in the case of Kerr-Sen-AdS$_{4}$ with $\tilde{a}=0$ and $\tilde{b}=1$. If $E\tilde{a}=\tilde{L}\Xi$, then $\tilde{r}=0$ is a double zero of $\tilde{R}$ for both the case of Kerr-AdS$_4$ \cite{Hackmann:2010zz} and the case of Kerr-Sen-AdS$_{4}$ with $\tilde{b}=0$. If $E\tilde{a}\neq \tilde{L}\Xi$ ($\tilde{b}\neq 0$), $\tilde{r}=0$ is a double zero of $\tilde{R}$ only for the case of Kerr-Sen-AdS$_{4}$.

\begin{figure}
	\begin{minipage}{0.4\textwidth}
    \includegraphics[scale=0.48,angle=0]{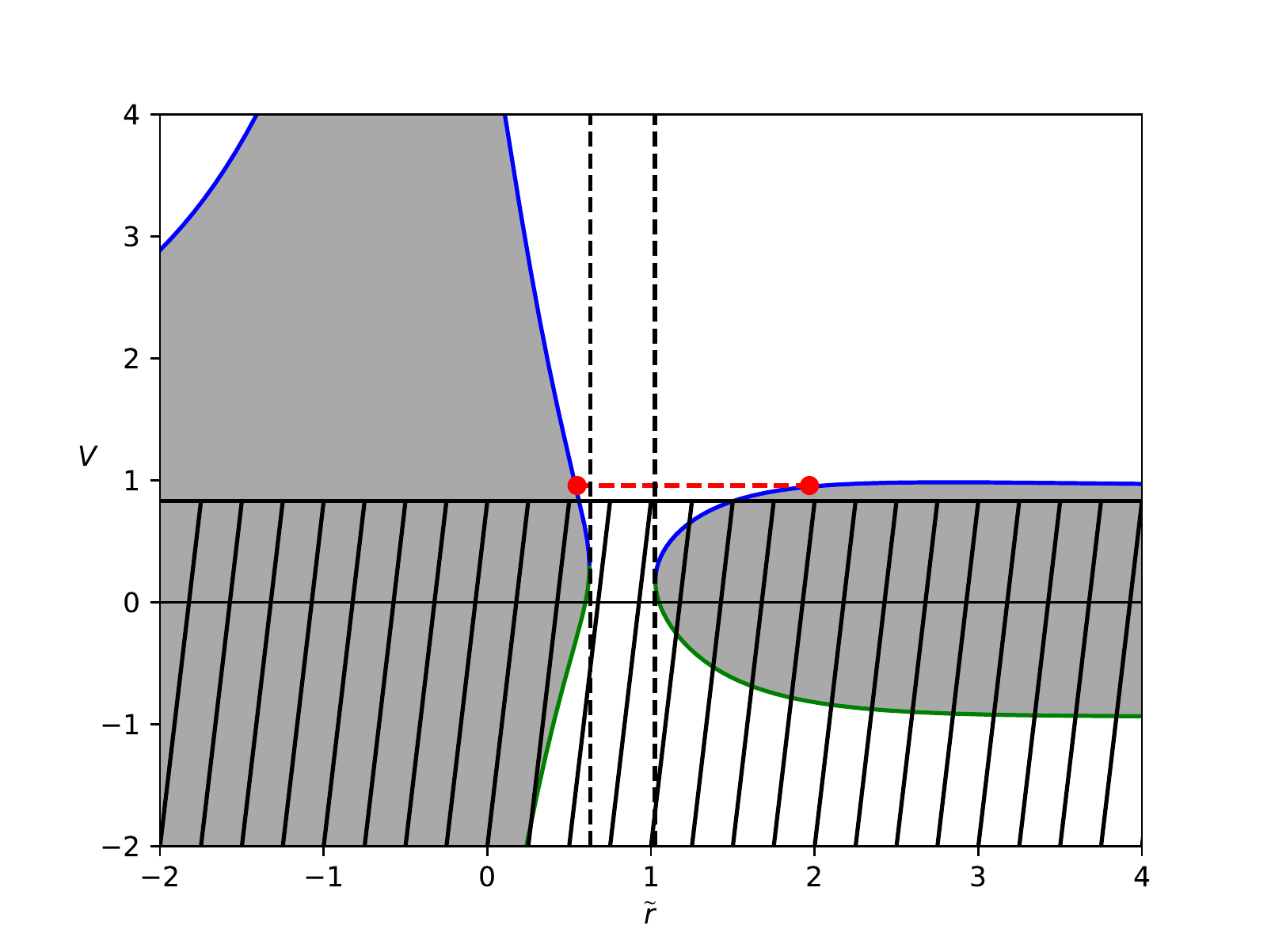}
	\end{minipage}
   \begin{minipage}{0.4\textwidth}
		\includegraphics[scale=0.48,angle=0]{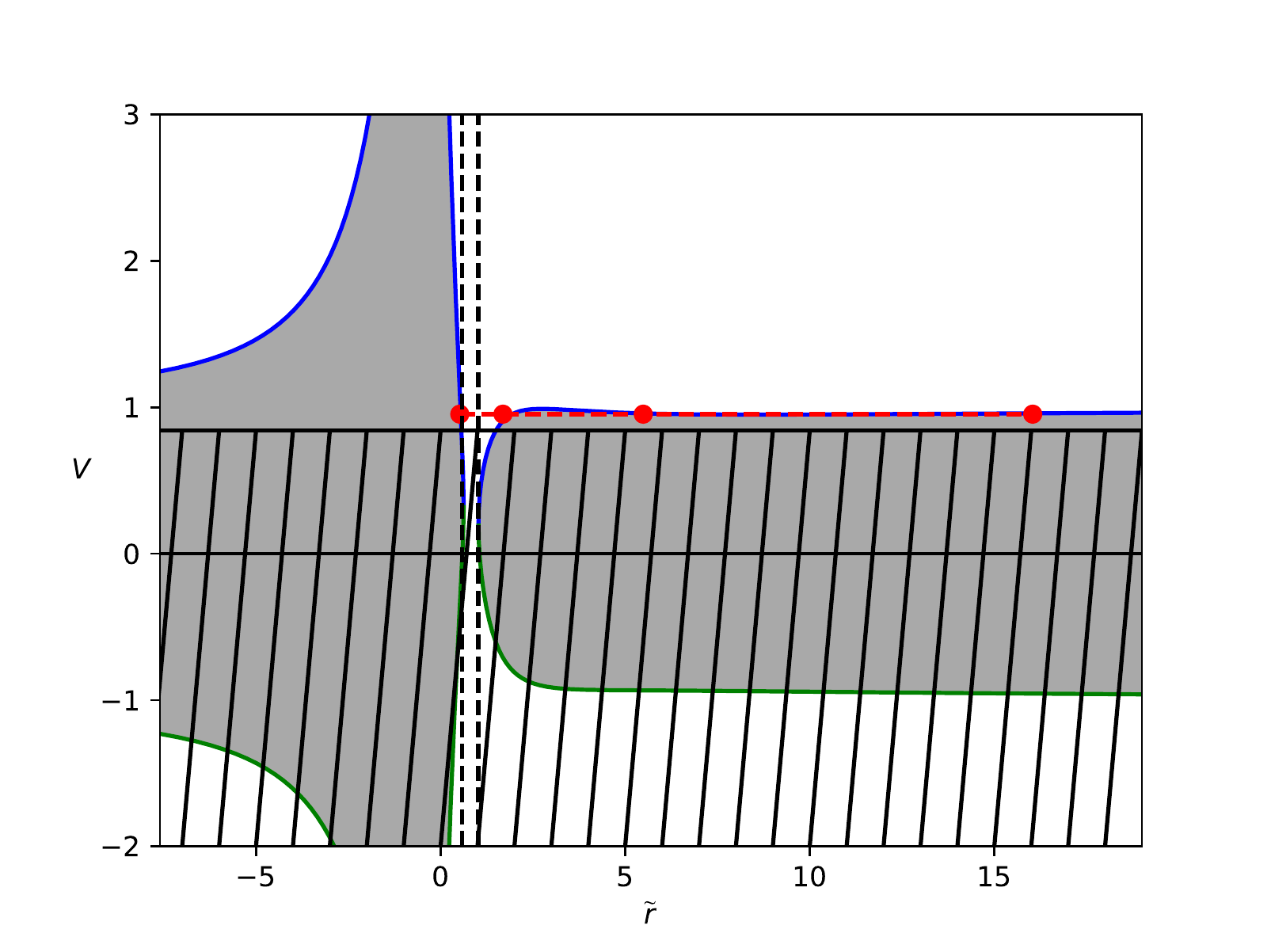}
	\end{minipage}
	\caption{Plots of the effective potential: $\varepsilon=1$, $\tilde{a}=0.8$, $\tilde{K}=12$, $\tilde{b}=0.175$, $\tilde{l}^{2}=\frac{1}{3}\times10^{-5}$, $\tilde{L}=0.45$ (for left column), and $\tilde{L}=0.5$ (for right column). The blue and green curves show the two branches of the effective potential. The red dashed lines correspond to energies. The red dots mark the zeros of the polynomial $R$. No motion is possible in the grey area. No $\theta$ geodesic motions are allowed in the oblique lines area dashed area. The vertical black dashed lines represent the position of the horizons.}
\label{potentia}
\end{figure}

In Fig. \ref{potentia}, we show the effective potential together with examples of energies for different orbit types. The green and blue
curves represent the two branches of the effective potential. The red dots which are the turning points of the orbits denote the zeros of the polynomial $\tilde{R}(\tilde{r})$. The red dashed lines in the grey area correspond to energies. Since $\tilde{R}(\tilde{r}) < 0$, no motion is possible in the grey area. The $\theta$ equation does not allow geodesic motion ($\tilde{\Theta}< 0$) in the oblique lines area.

\subsection{ISCO}
\begin{figure}
	\begin{minipage}{0.4\textwidth}
    \includegraphics[scale=0.48,angle=0]{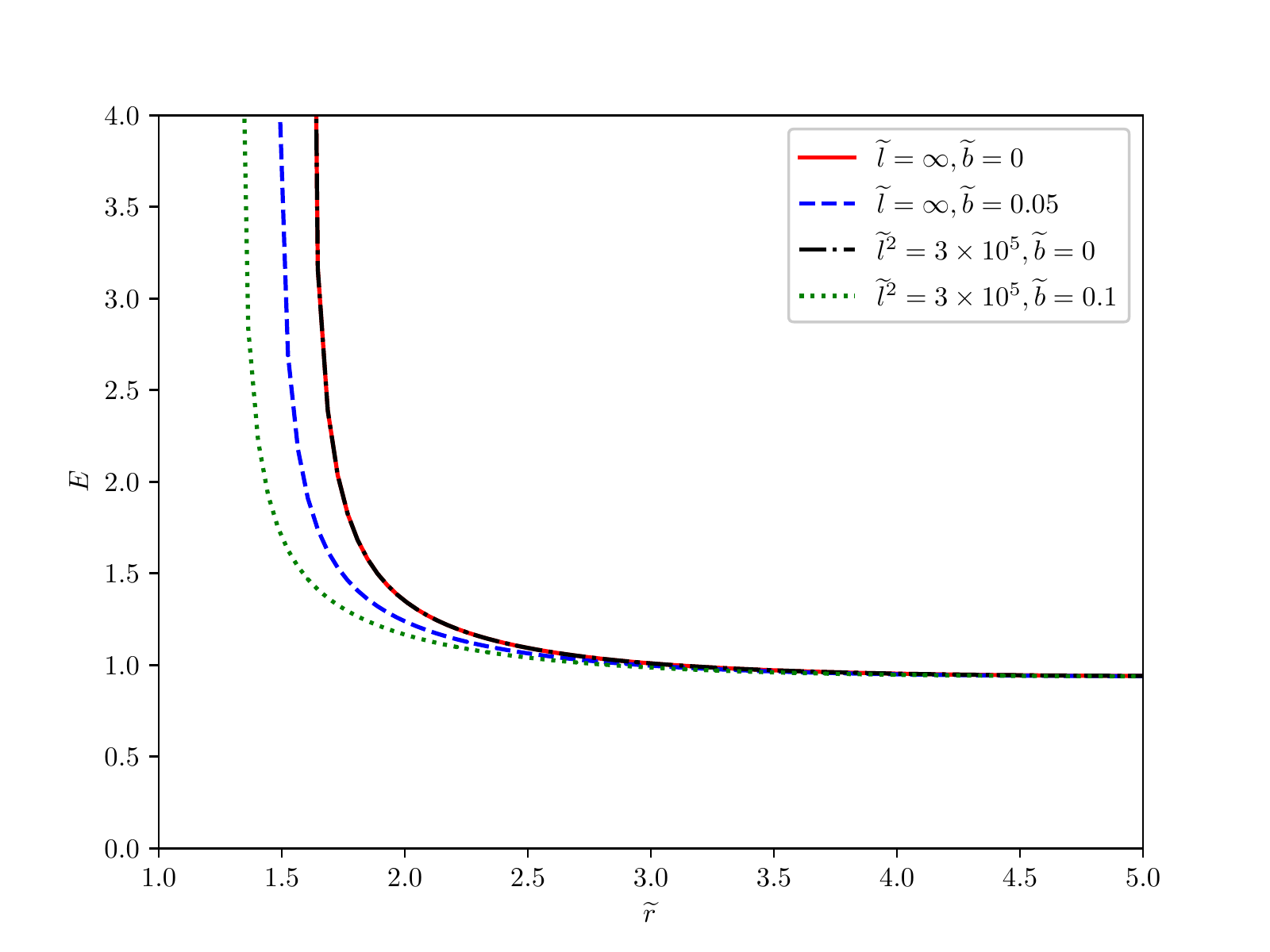}
	\end{minipage}
   \begin{minipage}{0.4\textwidth}
		\includegraphics[scale=0.48,angle=0]{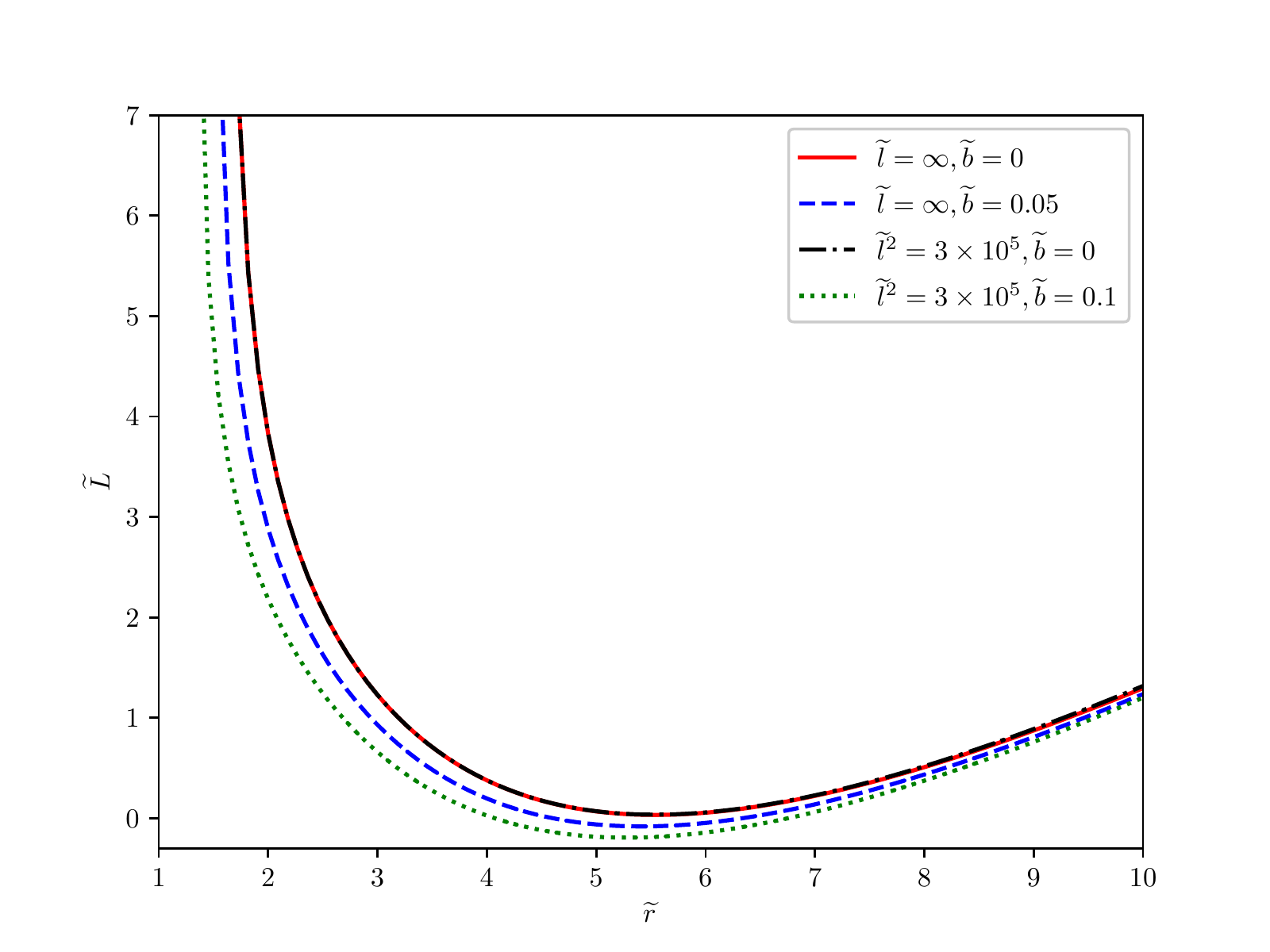}
	\end{minipage}
	\caption{$E^2-\tilde{r}$ and $\tilde{L}-\tilde{r}$ diagrams for some ISCOs with some special values of parameters.}
\label{ielr}
\end{figure}

For a test particle in the gravitational potential of a central
body, the innermost stable circular orbit (ISCO) is of importance as they represent the transition from stable orbit to those which fall through the event horizon. The ISCO is given by the conditions:
\begin{equation}
R(\tilde{r})=0, ~~~~\frac{dR}{d\tilde{r}}=0, ~~~~\frac{d^2R}{d\tilde{r}^2}=0.
\end{equation}
Namely, $r_{\rm{ISCO}}$ is the triple root of $R(\tilde{r})$. Stable spherical orbits with $\tilde{r}_0$ occur if radial coordinates adjacent to $\tilde{r}_0$ are not allowed due to $R(\tilde{r})$, which happens if $\tilde{r}_0$ is a maximum of $R$, namely, the orbit is
radially (vertically) stable (unstable) if $\partial^2_{\tilde{r}} \tilde{V}_{\rm{reff}}<0$ ($\partial^2_{\tilde{r}} \tilde{V}_{\rm{reff}}>0$). In Fig. \ref{ielr}, we plot $E^2-\tilde{r}$ and $\tilde{L}-\tilde{r}$ diagrams for some ISCOs in Kerr, Kerr-dilaton, Kerr-AdS$_{4}$, and Kerr-Sen-AdS$_{4}$ spacetime, respectively. We observe that $E^2$ decrease as $\tilde{r}$ increases, while $\tilde{L}$ decreases to some minimum value and then behave as a monotonously function. For large $\tilde{r}$, both $E^2$ and $\tilde{L}$ for there ISCOs run to the same values. But for small $\tilde{r}$, parameter $b$ will bring significant differences to $E^2$ and $\tilde{L}$, respectively.
\section{conclusion}
In this paper, we have discussed the motion of particles and light rays in the Kerr-Sen-AdS$_4$ spacetime. We have obtained the geodesic equations. Using the parametric diagrams, we have shown some regions where the $\tilde{r}$ and the $\theta$ geodesic motions are allowed. We have analysed in detail the impact of parameter related to dilatonic scalar on the orbit and found that it will result in more rich and complex orbit types, see for example, $\tilde{r}=0$ is not a double zero of $\tilde{R}$ in the case of Kerr-AdS$_4$ with $\tilde{a}=0$ \cite{Hackmann:2010zz}, but it is a double zero of $\tilde{R}$ in the case of Kerr-Sen-AdS$_{4}$ with $\tilde{a}=0$ and $\tilde{b}=1$. We also have discussed how the parameters of model affect the innermost stable circular orbit and show them with diagrams.

\begin{acknowledgments}
This study is supported in part by National Natural Science Foundation of China (Grant No. 12333008) and Hebei Provincial Natural Science Foundation of China (Grant No. A2021201034).
\end{acknowledgments}
\bibliographystyle{ieeetr}
\bibliography{geo}
\end{document}